\documentstyle[12pt,titlepage,epsf]{article}
\def\bbuildrel#1_#2^#3{\mathrel{\mathop{\kern 0pt#1}\limits_{#2}^{#3}}}
\def\slash#1{\setbox0=\hbox{$#1$}#1\hskip-\wd0\dimen0=5pt\advance
       \dimen0 by-\ht0\advance\dimen0 by\dp0\lower0.5\dimen0\hbox
         to\wd0{\hss\sl/\/\hss}}

\newcommand{\scs}{\scriptscriptstyle}
\newcommand{\be}{\begin{equation}}
\newcommand{\ee}{\end{equation}}
\newcommand{\bea}{\begin{eqnarray}}
\newcommand{\eea}{\end{eqnarray}}
\newcommand{\f}{\frac}

\newcommand{\al}{\alpha_s}
\newcommand{\e}{\epsilon}
\begin{document}
\begin{titlepage}

 \begin{flushright}
  {\bf TUM-HEP-336/98\\       
       IFT-14/98\\
      hep-ph/9901278\\
}
 \end{flushright}

 \begin{center}
  \vspace{0.6in}

\setlength {\baselineskip}{0.3in}
  {\bf \Large 
QCD corrections to FCNC decays mediated by $Z$-penguins and
$W$-boxes.}
\vspace{2cm} \\
\setlength {\baselineskip}{0.2in}

{\large  Miko{\l}aj Misiak$^{^{1,\star}}$
         and J{\"o}rg Urban$^{^{2}}$}\\

\vspace{0.2in}
$^{^{1}}${\it Physik Department, Technische Universit{\"a}t M{\"u}nchen,\\
                         D-85748 Garching, Germany}

\vspace{0.2in}
$^{^{2}}${\it Institut f{\"u}r Theoretische Physik, Technische Universit{\"a}t Dresden, \\
                        Mommsenstr. 13, D-01062 Dresden, Germany}

\vspace{2cm} 
{\bf Abstract } 
\end{center} 
\ \\
QCD corrections are evaluated to FCNC processes like $B \to X_s \nu
\bar{\nu}$, $K \to \pi \nu \bar{\nu}$, $B \to l^+ l^-$ or $K_L \to
\mu^+ \mu^-$, i.e. to processes mediated by effective operators
containing neutrino currents or axial leptonic currents. Such
operators originate from $W$-box and $Z$-penguin diagrams in the
Standard Model. QCD corrections to them are given by two-loop
diagrams. We confirm results for those diagrams which are already
present in the literature. However, our analytical expressions for the
Wilson coefficients disagree, due to a subtlety in regulating spurious
IR divergences. The numerical effect of the disagreement is rather
small. The size of the perturbative QCD corrections compared to the
leading terms depends on the renormalization scheme used at the
leading order. It varies from 0 to around 15\% for a reasonable class
of schemes. The uncertainty originating from uncalculated higher-order
(three-loop) QCD corrections is expected to be around 1\% .

\setlength{\baselineskip}{0.3in} 

\vspace{1cm}

\setlength {\baselineskip}{0.2in}
\noindent \underline{\hspace{2in}}\\ 
\noindent
$^\star$ {\footnotesize Address after January 1st, 1999: Theory Division,
  CERN, CH-1211 Geneva 23, Switzerland.\\ Permanent address: Institute
  of Theoretical Physics, Warsaw University, Ho\.za 69, 00-681 Warsaw, Poland.}

\end{titlepage} 

\setlength{\baselineskip}{0.3in}
\noindent {\bf 1. Introduction}

        Flavour Changing Neutral Current (FCNC) processes are
known to be an important source of information concerning the
Standard Model (SM) parameters, as well as a window towards new
physics.  However, predictions for their amplitudes are often
plagued with uncertainties due to low-energy QCD dynamics.
Processes dominated by $W$-box or $Z$-penguin diagrams which
involve both quarks and leptons form an important class of
exceptions from this rule. Low-energy dynamics can be then
efficiently factorized, and the theoretical predictions can be
made very precise for a given set of fundamental SM parameters.

        Phenomenologically, the most interesting processes
belonging to this class are 
$B \to X_s \nu \bar{\nu}$, 
$K_L \to \pi^0 \nu \bar{\nu}$,
$K^+ \to \pi^+ \nu \bar{\nu}$ 
and $B \to l^+ l^-$~ \cite{BB93}--\cite{tautau}. 
FCNC decays involving neutrinos in the final state are obviously
receiving contributions only from $W$-box or $Z$-penguin diagrams,
because the heavy bosons mediate all the neutrino interactions. In the
$B \to l^+ l^-$ case, photonic penguin contributions vanish due to the
electromagnetic current conservation for the lepton pair, while
double-photon intermediate states are numerically irrelevant.  The
analogous $K \to l^+ l^-$ decays are, on the contrary, dominated by
the CKM-favoured double-photon contributions and, in consequence, much
less theoretically clean.

\begin{figure}[htb] 
\centerline{
\epsfysize = 6cm
\epsffile{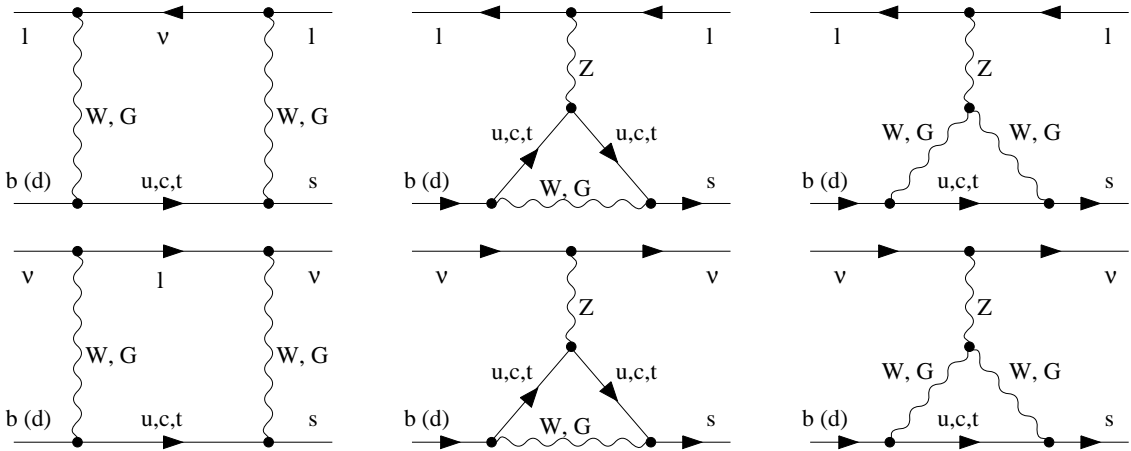}}
\caption{One-loop diagrams which give leading contributions
to the considered processes in the Standard Model. The charged
would-be Goldstone boson is denoted by $G$.}
\end{figure}

        The $W$-box and $Z$-penguin diagrams which give rise to
all those processes are presented in fig.~1.  Since external
momenta in these diagrams are of order $m_B$ or $m_K$, i.e. much
smaller than $M_W$, $M_Z$ or $m_t$, one can describe the
processes in question by introducing an effective theory which
is obtained from the SM by decoupling the heavy electroweak
bosons and the top quark
\be \label{Leff}
{\cal L}_{eff} = {\cal L}_{\scs QCD \times QED}\mbox{
(leptons and light quarks)} 
~~+~~ \f{4 G_F}{\sqrt{2}} \sum_n  C_n Q_n.
\ee
Here, $Q_n$ stand for ~dimension $>4$ effective interactions like
\bea 
\label{q1}
Q_1 &=&  \f{g_2^2}{8\pi^2} V^*_{ts} V_{tb} 
(\bar{s} \gamma_{\alpha} P_L b)(\bar{l} \gamma^{\alpha} P_L l),\\
\label{q2}
Q_2 &=& -\f{g_2^2}{8\pi^2} V^*_{ts} V_{tb} 
(\bar{s} \gamma_{\alpha} P_L b)(\bar{\nu} \gamma^{\alpha} P_L \nu),
\eea
where $g_2$ is the $SU(2)$ gauge coupling constant, $V_{ij}$ are the
CKM matrix elements, and $P_L = (1-\gamma_5)/2$. Numerical values of
their Wilson coefficients
\be
C_n = C_n^{(0)} + \f{\al}{4\pi} C_n^{(1)} + ... 
\ee
are found by matching the full and the effective theory Green's
functions. The matching is performed perturbatively in gauge
couplings and in (external momenta)$/M_W$.

        The two operators $Q_1$ and $Q_2$ written explicitly
above are the ones which give dominant contributions to 
$B \to l^+ l^-$ and $B \to X_s \nu \bar{\nu}$, respectively.
Operators relevant for $K^+ \to \pi^+ \nu \bar{\nu}$, $K_L
\to \pi^0 \nu \bar{\nu}$ and $K_L \to \mu^+ \mu^-$ are obtained 
from $Q_1$ and $Q_2$ by simply replacing $b$-quarks by $d$-quarks in
the operators themselves and in indices of the CKM factors.

        There are two main reasons for introducing the effective
theory description. One of them is summing up large QCD logarithms
like $[\al \ln (M_W^2/m_b^2)]^k$ from all orders of the SM
perturbation series. This is achieved by applying
renormalization group equations to the Wilson coefficients
$C_n$. The other reason is the fact, that different-looking
Feynman diagrams can give rise to very similar effective
operators whose nonperturbative matrix elements can be related
to each other.  Such relations are used e.g. for extracting
nonperturbative matrix elements in $K \to \pi \nu \bar{\nu}$
from the measurement of $K^+ \to \pi^0 \nu_e e^+$~\cite{MP96}.

        In the processes under consideration here, renormalization
group evolution of the Wilson coefficients plays only a minor role.
The dominant contributions originate from operators with only a single
(V--A) quark current, like those in eqns.~(\ref{q1}) and
(\ref{q2}). Such a current is not renormalized in a mass-independent
scheme, because it is conserved in the limit of vanishing quark
masses.  Consequently, anomalous dimensions of these operators vanish,
and RGE running of their Wilson coefficients may arise only at higher
orders in ~(light masses)$^2/M_W^2$.~ Such higher-order running is
numerically relevant (though subdominant) \linebreak \newpage
\noindent for $K^+ \to \pi^+ \nu \bar{\nu}$ and $(K_L \to l^+
l^-)_{\footnotesize Short~Distance}$ only.\footnote{
It arises from diagrams with double effective operator
insertions and charm-quark loops. They correspond to $W$-boxes
and $Z$-penguins with charm quarks on the full Standard Model
side. Such diagrams are suppressed by $m_c^2/M_W^2$ but 
enhanced by CKM angles with respect to the leading top-quark
contributions \cite{BB94}.}

        So long as the RGE running is negligible, and the
low-energy matrix elements can be found reliably, the main
calculational effort is required at the point of perturbative
matching of the full SM and the effective theory. Since masses
of the light particles can be set to zero when matching is
performed, Wilson coefficients of $Q_1$ and $Q_2$ are the same
as the Wilson coefficients of their $d$-quark analogs.  If the
Wilson coefficients of $Q_1$ and $Q_2$ were found only at the
leading order in QCD, i.e. from the purely electroweak one-loop
diagrams in fig.~1, they would contain an inherent uncertainty of
order 10\% due to two-loop gluonic corrections.  If those
two-loop corrections were not calculated, this uncertainty would
often become the main theoretical uncertainty in the final
predictions for branching ratios, for given values of the SM
parameters.

        A two-loop calculation of the next-to-leading matching
conditions for $Q_1$ and $Q_2$ was completed several years
ago \cite{BB93}. It has not been checked by any other group so
far.  In the present paper, we recalculate all the
next-to-leading contributions to the Wilson coefficients of
$Q_1$ and $Q_2$.  We confirm the results of ref.~\cite{BB93} for
all the necessary two-loop Feynman diagrams. However, our final
expressions for the Wilson coefficients disagree, due to a
subtlety in regulating spurious IR divergences.

        Our paper is organized as follows: In the next two sections,
we present a detailed description of evaluating Wilson
coefficients of $Q_1$ and $Q_2$. Section 2 is devoted to
the leading-order (one-loop) matching. The QCD corrections which
arise at two loops are taken into account in section 3.  Section
4 contains a discussion of numerical significance of the QCD
corrections.
        
\ \\
{\bf 2. Leading-order matching for $Q_1$ and $Q_2$.}

        Leading-order contributions to the Wilson coefficients
$C_1$ and $C_2$ can be found by requiring equality of
perturbative off-shell amplitudes generated by the full Standard
Model and the effective theory. We need to consider  
$b \to s l^+ l^-$ and $b \to s \nu \bar{\nu}$ transitions in the
cases of $Q_1$ and $Q_2$, respectively. On the Standard Model
side, we have to calculate the one-loop diagrams from fig.~1 for
vanishing external momenta. On the effective theory side, one
needs to include only tree-level diagrams with four-fermion
vertices corresponding to either $Q_1$ or $Q_2$.

        For purposes of the next section, we shall perform the
matching in $D = 4 - 2\e$ dimensions, i.e. without using
any identity from only 4-dimensional Dirac algebra. In
consequence, two other operators need to be introduced
explicitly on the effective theory side 
\bea 
\label{q1e}
Q_1^E &=&  \f{g_2^2}{8\pi^2} V^*_{ts} V_{tb} 
(\bar{s} 
\gamma_{\alpha_1} \gamma_{\alpha_2} \gamma_{\alpha_3} 
P_L b)(\bar{l} 
\gamma^{\alpha_3} \gamma^{\alpha_2} \gamma^{\alpha_1} 
P_L l)  \;\; - \;\; 4 Q_1,\\
\label{q2e}
Q_2^E &=& -\f{g_2^2}{8\pi^2} V^*_{ts} V_{tb} 
(\bar{s} 
\gamma_{\alpha_1} \gamma_{\alpha_2} \gamma_{\alpha_3} 
P_L b)(\bar{\nu} 
\gamma^{\alpha_1} \gamma^{\alpha_2} \gamma^{\alpha_3} 
P_L \nu) \;\; - \;\; 16 Q_2.
\eea
The superscript "E" stands for the name "evanescent" which
originates from the fact that such operators vanish in 4
dimensions due to the identity 
\be
\gamma_{\alpha_1} \gamma_{\alpha_2} \gamma_{\alpha_3} =
  g_{\alpha_1 \alpha_2} \gamma_{\alpha_3}
- g_{\alpha_1 \alpha_3} \gamma_{\alpha_2}
+ g_{\alpha_2 \alpha_3} \gamma_{\alpha_1}
+ i \varepsilon_{\beta \alpha_1 \alpha_2 \alpha_3 } \gamma^{\beta} \gamma_5.
\ee
This identity cannot be analytically extended to D dimensions.
Operators $Q^E_1$ and $Q^E_2$ arise in D-dimensional matching, because
$W$-box diagrams in fig.~1 contain triple products of Dirac matrices
which cannot be reduced to anything simpler in D-dimensions.

        Elementary calculation of the one-loop diagrams in
fig.~1 with vanishing external momenta in the Feynman-'t~Hooft
gauge gives us the following results for the leading-order
contributions to the considered Wilson coefficients\footnote{ 
Here, we ignore contributions suppressed by $m_c^2/M_W^2$. They
are relevant in $K^+ \to \pi^+ \nu \bar{\nu}$, i.e. for the
$d$-quark analog of $Q_2$, because of their CKM enhancement.} 
\bea
C_1^{(0)}  &=& C_0(x_t)   -           B_0(x_t) + {\cal O}(\e),\\
C_2^{(0)}  &=& C_0(x_t)   -         4 B_0(x_t) + {\cal O}(\e),\\
C_1^{E(0)} &=& C_2^{E(0)} = -\f{1}{4} B_0(x_t) + {\cal O}(\e),
\eea
where $x_t = m_t^2/M_W^2$ and \cite{IL81}
\bea
B_0(x) = \f{      x}{4(x-1)^2} \ln x - \f{     x}{4(x-1)},\\
C_0(x) = \f{3x^2+2x}{8(x-1)^2} \ln x + \f{x^2-6x}{8(x-1)}.
\eea
The functions $B_0$ and $C_0$ originate from $W$-box and
$Z$-penguin diagrams, respectively.

\ \\
{\bf 3. Next-to-leading order matching for $Q_1$ and $Q_2$.}

        Before performing the NLO matching, we need to learn the
structure of one-loop UV counterterms on the effective theory
side. Evaluating the UV QCD counterterms can be done separately in the
lepton $(Q_1,Q^E_1)$ and neutrino $(Q_2,Q^E_2)$ sectors, thanks to
their different leptonic content. In each of the two sectors, bare
quantities are replaced by the QCD-renormalized ones as follows:
\be
C Q  +  C^E Q^E  \to  Z_{\psi} \left( C   Z_{NN} Q + C   Z_{NE} Q^E 
                                    + C^E Z_{EN} Q + C^E Z_{EE} Q^E \right).
\ee
Here, $Z_{\psi}$ denotes the usual quark wave-function renormalization
constant. The remaining renormalization constants are found in
the MS scheme from the following two conditions which the
$n$-loop effective theory amplitudes have to satisfy \cite{evan}: 
\begin{itemize}
\item{} Renormalized amplitudes proportional to the coefficient $C$ of
the "normal" operator $Q$ have to be finite in the limit $D \to
4$. Counterterms which make them finite can contain nothing but
$1/\e^k$ poles, with $1 \leq k \leq n$.
\item{} Renormalized amplitudes proportional to the coefficient
$C^E$ of the "evanescent" operator $Q^E$ have to {\em vanish} in
the limit $D \to 4$. Counterterms which make them
vanish can contain nothing but $1/\e^k$ poles, with $0 \leq k
\leq n-1$ if they are proportional to the "normal" operator,
and with $1 \leq k \leq n$ if they are proportional to the 
"evanescent" operator. This means that uniquely defined finite
counterterms occur in this case, too.
\end{itemize}

\begin{figure}[htb] 
\centerline{
\epsfysize = 3cm
\epsffile{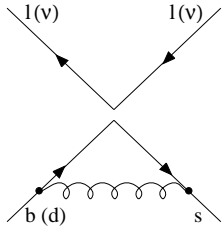}}
\caption{One-loop diagram which determines renormalization constants
in the effective theory.}
\end{figure}

        Recovering the one-loop renormalization constants in the
present case is extremely easy, because there is only a single
one-loop diagram to consider. It is shown in fig.~2. The
four-fermion vertex in this diagram is either the "normal"
operator $Q$ or the "evanescent" operator $Q^E$. Since at this
point we are interested only in the $1/\e$ UV poles originating
from loop momentum integrals, we set the external momenta and
the physical masses to zero, but keep a common mass parameter as the
IR regulator in all the propagator denominators \cite{CMM98}.
We find
\be \label{renor.const}
Z_{NN} = 1, \hspace{1cm}
Z_{NE} = 0, \hspace{1cm}
Z_{EN} = \f{\al}{4\pi} r + {\cal O}(\al^2), \hspace{1cm}
Z_{EE} = 1 + {\cal O}(\al^2),
\ee
with $r=32$ and $r=-32$ in the lepton $(Q_1,Q^E_1)$ and neutrino
$(Q_2,Q^E_2)$ sectors, respectively.

        The above equations mean that the renormalization constant
$Z_{\psi}$ was always enough to remove the $1/\e$ pole from the
one-loop diagram in fig.~2. Actually, the first two equations are true
to all orders in QCD, because the quark current in $Q$ is conserved
for vanishing quark masses, as already mentioned in the
introduction. The important part of eqn.~(\ref{renor.const}) is the
finite renormalization constant $Z_{EN}$ which is going to affect our
final results.

\begin{figure}[htb] 
\centerline{
\epsfysize = 2.5cm
\epsffile{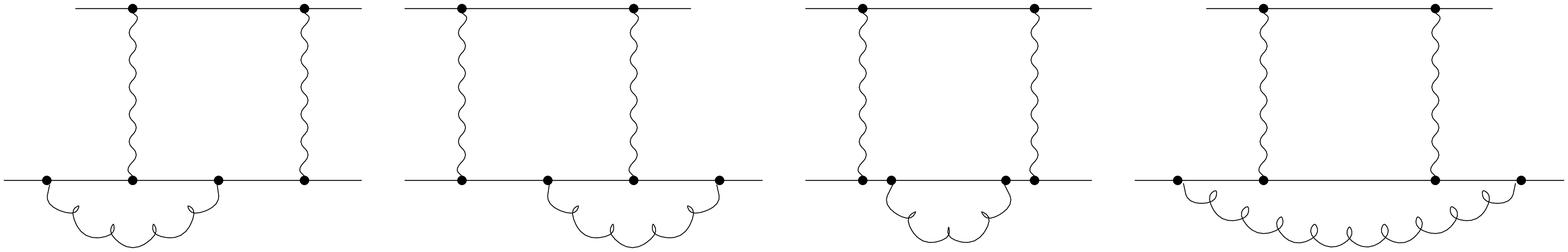}}
\caption{Two-loop QCD corrections to the $W$-box diagrams from fig.~1.}
\end{figure}
\begin{figure}[htb] 
\centerline{
\epsfysize = 7.5cm
\epsffile{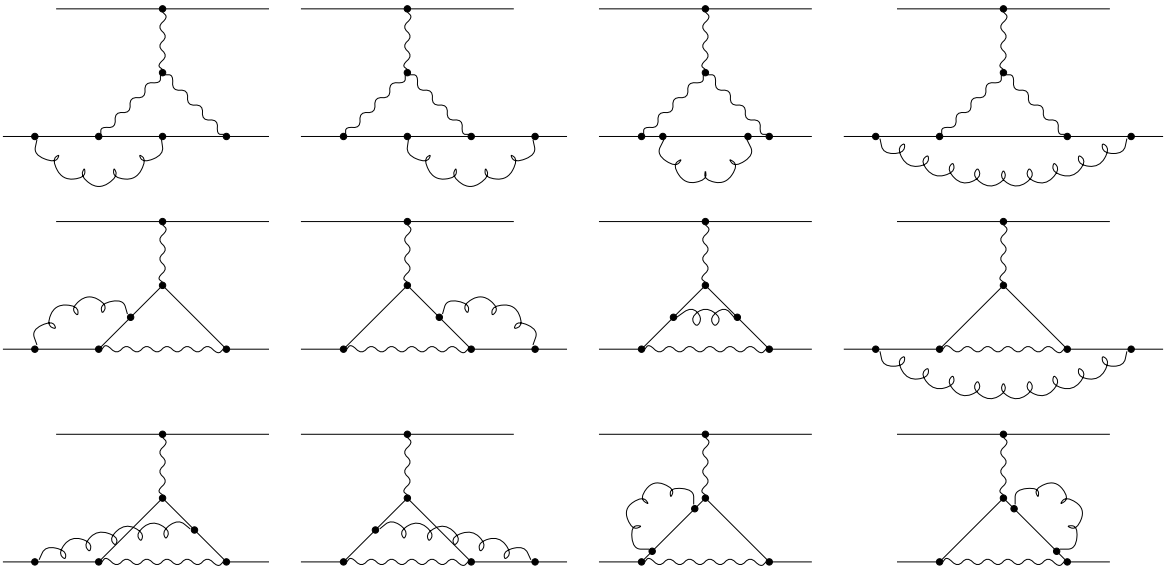}}
\caption{Two-loop QCD corrections to the $Z$-penguin diagrams from fig.~1.}
\end{figure}
\begin{figure}[htb] 
\centerline{
\epsfysize = 2cm
\epsffile{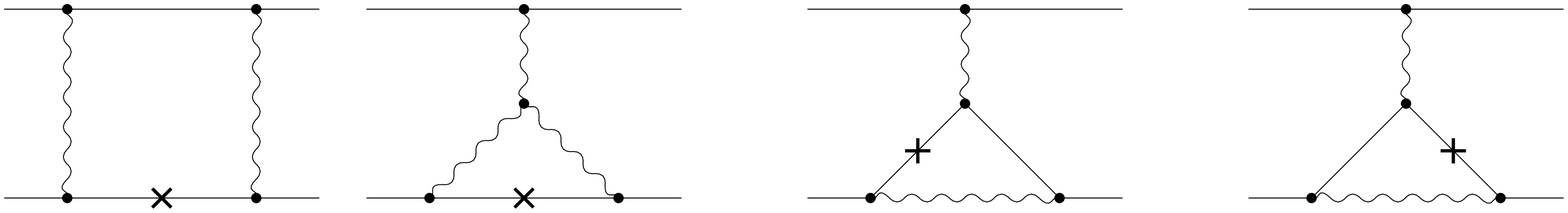}}
\caption{Diagrams with quark mass counterterms.}
\end{figure}

        We are now ready to perform the NLO matching, i.e. to find
$C_1^{(1)}$ and $C_2^{(1)}$. On the full Standard Model side, one
needs to calculate the two-loop diagrams shown in figs.~3 and 4, as
well as the diagrams with quark mass counterterms in fig.~5. We ignore
the quark wave-function renormalization now. This is allowed so long
as one does it on the effective theory side of the matching equation,
too. As in the previous section, it is enough to calculate all the
diagrams in the limit of vanishing external momenta. The calculation
is particularly simple when masses of all the light particles are set
to zero. Then, the 1PI parts of the diagrams depend on masses only via
the ratio $x_t$, and they are very easy to handle with recurrence
relations for two-loop vacuum integrals \cite{DT93}.

        Setting light masses to zero may lead to generation of
spurious IR divergences. They should cancel with similar
divergences on the effective theory side, so that the final
results for the Wilson coefficients are finite in the limit $D
\to 4$. This cancellation occurs only after comparing the full
and effective theory amplitudes. Thus, one needs to calculate
both amplitudes in D dimensions, even after their UV
renormalization. This was the reason why we needed to perform
the matching for $Q_k^E$ in the previous section.

        The sum of the diagrams in figs.~3, 4 and 5 turns out to be
finite in the limit $D \to 4$. However, as we shall see in the
following, it contains finite terms originating from multiplication of
$1/\e$ spurious IR divergences with ${\cal O}(\e)$ terms from the
Dirac algebra.\\

        Calculating the effective theory side of the
next-to-leading matching condition is particularly simple,
because we have set all the light masses to zero on the full
theory side.  Thus, we now need to do the same on the effective
theory side, where only light particles occur. Consequently, all
the loop diagrams on the effective theory side vanish in
dimensional regularization. What remains are only tree-level
diagrams with insertions of either the original operator
vertices or the UV counterterms. The matching equation takes
then the following form:
\be \label{match} \hspace*{-4mm}
\left( \hspace{-2mm} \begin{array}{c} \mbox{\scriptsize 1- and 2-loop}\\[-2mm]
        \mbox{\scriptsize SM diagrams} \end{array} \hspace{-2mm} \right)
= \mbox{(const.)} \times 
\left\{ \left[ C^{(0)} + \f{\al}{4 \pi} \left( C^{(1)} + r C^{E(0)} \right) \right]
\langle Q \rangle_{tree} \;+\; 
\left( \hspace{-2mm} \begin{array}{c} \mbox{\scriptsize something}\\[-2mm]
                   \mbox{\scriptsize finite in~} 
                         {\scriptstyle D=4} \end{array} \hspace{-2mm} \right) 
\langle Q^E \rangle_{tree}  \right\}.
\ee
We have now consistently ignored the quark wave-function renormalization
on both the full and effective theory sides.

        Vanishing of all the loop diagrams on the effective
theory side means, in particular, that all the UV divergences in
those diagrams cancel with spurious IR divergences (which arise
due to setting all the external momenta and masses to zero).
Thus, UV counterterms on the effective theory side actually
reproduce the spurious IR divergences. The spurious IR
divergences are the same on the full and effective theory sides,
and cancel in the matching equation. In our case, there are no
$1/\e$ poles involved in this cancellation. We only have the finite
"$r C^{E(0)}$" term.  However, $1/\e$ poles would occur in a
generic case.\footnote{ 
They do occur e.g. when two-loop off-shell
matching for $b \to s \gamma$ is performed for vanishing light
particle masses.} 
Thus, the "$r C^{E(0)}$" term on the effective theory side can
be interpreted as the one which cancels finite terms originating
from multiplication of $1/\e$ spurious IR divergences with
${\cal O}(\e)$ terms from the Dirac algebra on the full theory
side.

        Since we do not need to recover the NLO Wilson coefficients
of the evanescent operators, we can take the limit $D \to 4$ in the
matching equation (\ref{match}). Then $\langle Q^E \rangle_{tree}$
vanishes, and we find
\bea 
\label{c11}
C_1^{(1)}  = C_1(x_t)   -     B_1(x_t,-1/2) + {\cal O}(\e),\\
\label{c21}
C_2^{(1)}  = C_1(x_t)   -   4 B_1(x_t,+1/2) + {\cal O}(\e),
\eea
where
\bea 
\label{c1x}
C_1(x) &=& 
  \f{x^3+4x}{(x-1)^2} Li_2(1-x) 
+ \f{x^4-x^3+20x^2}{2(x-1)^3} \ln^2 x
+ \f{-3x^4-3x^3-35x^2+x}{3(x-1)^3} \ln x
\nonumber \\[2mm] &&
+ \f{4x^3+7x^2+29x}{3(x-1)^2}
+ 8 x \f{\partial C_0(x)}{\partial x} \ln \f{\mu^2}{M_W^2},\\[2mm]
\label{b1mx}
B_1(x,-1/2) &=& 
  \f{2x}{(x-1)^2} Li_2(1-x) 
+ \f{3x^2+x}{(x-1)^3} \ln^2 x
+ \f{-11x^2-5x}{3(x-1)^3} \ln x 
\nonumber \\[2mm] &&
+ \f{-3x^2+19x}{3(x-1)^2} 
+ 8 x \f{\partial B_0(x)}{\partial x} \ln \f{\mu^2}{M_W^2},\\[2mm]
\label{b1px}
B_1(x,+1/2) &=& 
  \f{2x}{(x-1)^2} Li_2(1-x) 
+ \f{3x^2+x}{(x-1)^3} \ln^2 x
+ \f{-31x^2-x}{6(x-1)^3} \ln x 
\nonumber \\[2mm] &&
+ \f{3x^2+29x}{6(x-1)^2}
+ 8 x \f{\partial B_0(x)}{\partial x} \ln \f{\mu^2}{M_W^2}, 
\eea
Here, $\mu$ stands for the renormalization scale at which the
top quark mass is $\overline{MS}$-renormalized. The functions
$B_1(x_t,\pm 1/2)$ and $C_1(x_t)$ originate from gluonic corrections
to $W$-boxes and $Z$-penguins, respectively. The difference
between $B_1(x,-1/2)$ and $B_1(x,+1/2)$ is proportional
to $B_0(x)$
\be
B_1(x,-1/2) - B_1(x,+1/2) = 6 B_0(x).
\ee

        Our result for the function $C_1(x)$ is in perfect
agreement with ref.~\cite{BB93}. However, the results for
$B_1(x,\pm 1/2)$ disagree. The disagreement is due to the term
"$r C^{E(0)}$" in our matching equation (\ref{match}). 

        Since our results differ from the previously published ones
due to a subtlety in regulating spurious IR divergences, it is
worthwhile to cross-check them using light quark masses as IR
regulators. Calculation of the 2-loop SM diagrams is then somewhat
more complicated. One-loop diagrams on the effective theory side are
nonvanishing in this case, and one needs to calculate their finite
parts.  However, the evanescent operators are unimportant. The
matching equation can be written in 4 dimensions. Both the effective
and the full theory amplitudes depend on the light quark masses in the
same way. Terms dependent on the light quark masses cancel out in the
matching equation. The Wilson coefficients we obtain in the end are
exactly the same as the ones already found in
eqns.~(\ref{c11})--(\ref{b1px}). Thus, our NLO matching results are
cross-checked by calculating them twice with use of two different
regulators for spurious IR divergences: dimensional regularization and
light quark masses.

\ \\
{\bf 4. Numerical significance of the NLO corrections}

        Branching ratios of the decays $B \to X_s \nu \bar{\nu}$ and
$K_L \to \pi^0 \nu \bar{\nu}$ are proportional to the squared Wilson
coefficient $|C_2|^2$, while $B \to l^+ l^-$ is proportional to
$|C_1|^2$. Thus, numerical significance of the NLO matching in these
decays can be seen in figs.~6 and 7 where $|C_k|^2$ are plotted as
functions of the matching scale $\mu$.  Dependence on this scale
enters via \linebreak $x_t = [m_t^{\overline{MS}}(\mu)/M_W]^2$,~
$\al(\mu)$ and explicitly in $C_k^{(1)}$.  We use $m_t^{pole}=175$~GeV
and $\al(M_Z) = 0.118$.  Solid lines present the full NLO predictions,
while dashed lines correspond to the LO results. Dotted lines indicate
what the NLO results would be if the "$r C^{E(0)}$" term was not
included in the matching equation.

\begin{figure}[htb]
\vspace{-4cm} 
\centerline{
\epsfysize = 15cm
\epsffile{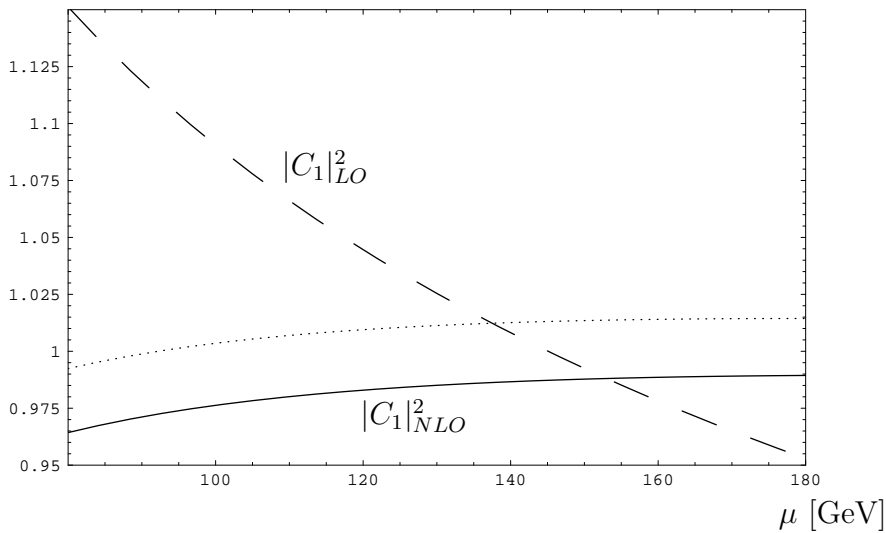}}
\vspace{-9cm}
\centerline{ \hspace{-2.5cm} $|C_1|^2_{LO}$} 
\vspace{2.8cm}
\centerline{ $|C_1|^2_{NLO}$} 
\vspace{0.8cm}
\centerline{ \hspace{11cm} $\mu$~[GeV]} 
\caption{$|C_1(\mu)|^2$ at the leading and next-to-leading order.}
\end{figure}
\begin{figure}[htb] 
\vspace{-4cm} 
\centerline{
\epsfysize = 15cm
\epsffile{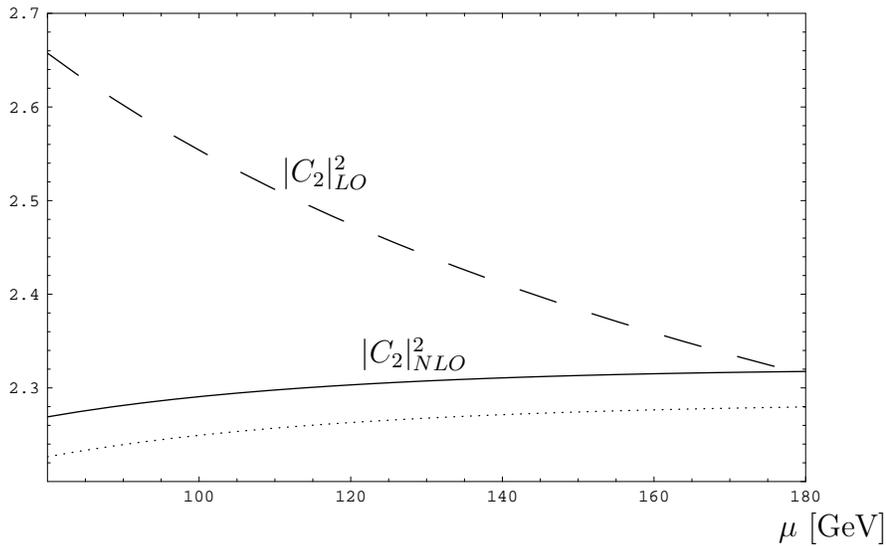}}
\vspace{-9cm}
\centerline{ \hspace{-2.5cm} $|C_2|^2_{LO}$} 
\vspace{1.9cm}
\centerline{ $|C_2|^2_{NLO}$} 
\vspace{1.8cm}
\centerline{ \hspace{11cm} $\mu$~[GeV]} 
\caption{$|C_2(\mu)|^2$ at the leading and next-to-leading order.}
\end{figure}

        One can see that the NLO QCD corrections to $|C_k|^2$ range
from 0 to around 15\% for a reasonable class of renormalization
schemes, i.e. for $\mu \in [M_W, m_t]$. The remaining $\mu$-dependence
at NLO is an estimate of uncalculated higher order (three-loop) QCD
corrections. It is around 1\%.  The numerical effect of the "$r
C^{E(0)}$" term is seen to be rather small (around 2\%).  To some
extent, it is due to the fact that it originates from
$W$-boxes. Contributions from $W$-boxes tend to a constant at large
$x_t$, while $Z$-penguin contributions grow with $x_t$, and are
dominant for the physical value of $x_t \simeq 5$. Thus, from the
numerical point of view, our results are a confirmation of what has
been found in ref.~\cite{BB93} and in the following phenomenological
analyses \cite{AB.phen}.

        The decays $K^+ \to \pi^+ \nu \bar{\nu}$ and $K_L \to \mu^+
\mu^-$ are somewhat more complicated because their amplitudes
receive other significant contributions which are not proportional to
the Wilson coefficients discussed in sections 2 and 3. Both of them
are affected by ${\cal O}(m_c^2/M_W^2)$ terms which are CKM-enhanced
with respect to the leading top-quark contributions. Those charm-quark
loops are numerically as large as 20\%-40\% of the top-quark ones
\cite{BB94}. In addition, the $K_L \to \mu^+ \mu^-$ mode is actually
dominated by contributions from double-photon intermediate states
which arise at higher order in $\alpha_{em}$, but involve u-quark
loops and, in consequence, are strongly CKM enhanced with respect to
the $W$-box and $Z$-penguin diagrams. Thus, QCD corrections to the
top-quark loops are of mild importance for $K^+ \to \pi^+ \nu
\bar{\nu}$, and totally unimportant in the $K_L \to \mu^+ \mu^-$
case.

\ \\
{\bf 5. Summary}
        
        We have evaluated QCD corrections to processes mediated
by $Z$-penguin and $W$-box diagrams. Two-loop matching of the
full Standard Model and the effective theory was performed 
off-shell, for vanishing external momenta and light particle
masses. In consequence, calculation of the necessary two-loop
Feynman diagrams was very easy, but the matching had to be
performed in D dimensions, due to dimensionally regulated
spurious IR divergences. In effect, Wilson coefficients of
evanescent operators had to be found, and they affected the
final physical results. Such a phenomenon is expected to take
place in a generic process of heavy particle decoupling, when
light particle masses are set to zero in the matching procedure.

        We have confirmed the results of ref.~\cite{BB93} for the
two-loop diagrams, but our final expressions for the NLO Wilson
coefficients disagree due to the very contribution from the evanescent
operators. We have cross-checked our results using light quark masses
as IR regulators, in which case there was no need to consider
evanescent operators.

        The effect of the NLO QCD corrections on the branching ratios
varies between 0 and around 15\% in a reasonable class of
renormalization schemes. Higher order effects are expected to be
around 1\%. The effect of the evanescent operator contribution in the
matching turns out to be small. Thus, from the numerical point of
view, our results are a confirmation of what has been found in
ref.~\cite{BB93} and in the following phenomenological analyses
\cite{AB.phen}.

\ \\
{\bf  Acknowledgments}

        We thank Andrzej Buras and Gerhard Buchalla for useful
correspondence, for positive verification of our results and for
carefully reading the manuscript of the present paper. J.U. thanks
C.~Bobeth, K.~Schubert, R.~Sch\"utzhold and G.~Soff for helpful
discussions. J.U. and M.M. have been supported in part by the German
Bundesministerium f{\"u}r Bildung und Forschung under contracts
06~DD~823 and 06~TM~874, respectively. M.M. has been supported in part
by the DFG project Li~519/2-2, as well as by the Polish Committee for
Scientific Research under grant 2~P03B~014~14, 1998-2000.
        
\newpage
\setlength {\baselineskip}{0.2in}


\begin{thebibliography}{99}
\newcommand{\np}[3]{Nucl. Phys. {\bf B#1} (19#2) #3}
\newcommand{\pl}[3]{Phys. Lett. {\bf B#1} (19#2) #3}
\newcommand{\pr}[3]{Phys. Rev.  {\bf D#1} (19#2) #3}
\newcommand{\prl}[3]{Phys. Rev. Lett. {\bf #1} (19#2) #3}
\newcommand{\prp}[3]{Phys. Rept. {\bf #1} (19#2) #3}
\newcommand{\rmp}[3]{Rev. Mod. Phys. {\bf #1} (19#2) #3}
\newcommand{\zpc}[3]{Z. Phys. {\bf C#1} (19#2) #3}

\bibitem{BB93} G.~Buchalla and A.J.~Buras, \np{398}{93}{285},
               \np{400}{93}{225}.
\bibitem{BB94} G.~Buchalla and A.J~Buras, \np{412}{94}{106}\\
               and preprint TUM-T31-337/98.
\bibitem{rev1} G.~Buchalla, A.J~Buras and M.~Lautenbacher, 
               \rmp{68}{96}{1125}.
\bibitem{rev2} A.J.~Buras and R.~Fleischer, {\it Quark mixing, 
               CP-violation and rare decays after the top quark 
               discovery}, in the Review Volume {\it Heavy Flavors II},  
               eds. A.J.~Buras and M.~Lindner, Advanced Series on Directions
               in High Energy Physics, World Scientific Publishing Co., 
               Singapore, 1998~ (hep-ph/9704376).
\bibitem{rev3} A.J.~Buras, {\it Weak Hamiltonian, CP violation and
               rare decays}, in {\it Probing the Standard Model of
               particle interactions}, eds. F.~David and B.~Gupta,
               Elsevier Science B.V., 1998~ (hep-ph/9806471).
\bibitem{tautau} Y. Grossman, Z.~Ligeti and E.~Nardi, \pr{55}{97}{2768}.
\bibitem{MP96} W.~Marciano and Z.~Parsa, \pr{53}{96}{1}.
\bibitem{IL81} T.~Inami and C.S.~Lim, Progr. Theor. Phys. {\bf 65} (1981) 297.
\bibitem{DT93} A.I.~Davydychev and J.B.~Tausk, \np{397}{93}{123}.
\bibitem{evan} A.J.~Buras and P.H.~Weisz, \np{333}{90}{66};\\
               M.~J.~Dugan and B.~Grinstein, \pl{256}{91}{239};\\
               S.~Herrlich and U.~Nierste, \np{455}{95}{39}.
\bibitem{CMM98} K.~Chetyrkin, M.~Misiak and M.~M\"unz, \np{518}{98}{473}.
\bibitem{AB.phen} G.~Buchalla and A.J.~Buras, \pl{333}{94}{221};\\
                                  A.J.~Buras, \pl{333}{94}{476};\\
                  G.~Buchalla and A.J.~Buras, \pl{336}{94}{263};\\
                  G.~Buchalla and A.J.~Buras, \pr{54}{96}{6782}.\\
\end{thebibliography}
\end{document}